\documentclass[aps,reprint,twocolumn,superscriptaddress,10pt]{revtex4-1}
\usepackage{siunitx}
\usepackage{amsfonts}
\usepackage{amsmath}
\usepackage{amssymb}
\usepackage{graphicx}
\usepackage{dcolumn}
\usepackage{mciteplus}
\usepackage{euscript}
\usepackage{multirow}
\usepackage{color}
\usepackage{siunitx}
\providecommand{\U}[1]{\protect\rule{.1in}{.1in}}
\graphicspath{{./figs/}}

\providecommand{\U}[1]{\protect\rule{.1in}{.1in}}
\newcommand{\redtx}[1]{\textcolor{red}{}}

\newcommand{\uu}{\uparrow\uparrow}
\newcommand{\ud}{\uparrow\downarrow}
\newcommand{\du}{\downarrow\uparrow}
\newcommand{\dd}{\downarrow\downarrow}

\newcommand{\tmalb}{$T_{2}$AlB$_{2}$}
\newcommand{\cralb}{Cr$_{2}$AlB$_{2}$}
\newcommand{\mnalb}{Mn$_{2}$AlB$_{2}$}
\newcommand{\fealb}{Fe$_{2}$AlB$_{2}$}
\newcommand{\coalb}{Co$_{2}$AlB$_{2}$}
\newcommand{\nialb}{Ni$_{2}$AlB$_{2}$}
\newcommand{\fecralb}{(Fe$_{1-x}$Cr$_{x}$)$_2$AlB$_{2}$}
\newcommand{\femnalb}{(Fe$_{1-x}$Mn$_{x}$)$_2$AlB$_{2}$}
\newcommand{\fecoalb}{(Fe$_{1-x}$Co$_{x}$)$_2$AlB$_{2}$}
\newcommand{\fenialb}{(Fe$_{1-x}$Ni$_{x}$)$_2$AlB$_{2}$}
\newcommand{\fetaalb}{Fe$_{2-x}T_{x}$AlB$_{2}$}
\newcommand{\fetbalb}{(Fe$_{1-x}T_{x}$)$_2$AlB$_{2}$}
\newcommand{\fealbc}{Fe$_2$Al(B$_{0.75}$C$_{0.25}$)$_2$}

\newcommand{\fealbz}{Fe$_2$Al(B$_{0.75}Z_{0.25}$)$_2$}

\newcommand{\fealzb}{Fe$_2$(Al$_{0.5}Z_{0.5}$)B$_2$}
\newcommand{\Cr}{\text{Cr}}
\newcommand{\Mn}{\text{Mn}}
\newcommand{\Fe}{\text{Fe}}
\newcommand{\Co}{\text{Co}}
\newcommand{\Ni}{\text{Ni}}
\newcommand{\NM}{\text{NM}}
\newcommand{\FM}{\text{FM}}
\newcommand{\AFM}{\text{AFM}}

\def\efermi{E_{\rm F}}
\def\mscell{M}
\def\mspin{m_{i}}
\def\jij{J_{ij}}
\def\tc{T_\text{C}}
\def\tn{T_\text{N}}
\def\etal{\textit{et al.}}
\def\abinitio{\emph{ab initio}}

\newcommand{\frfig}[1]{Figure ~\ref{#1}}
\newcommand{\rfig}[1]{Fig.~\ref{#1}}
\newcommand{\rtbl}[1]{Table~\ref{#1}}

\begin{document}

\title{Electronic structure and magnetic properties in $T_2$AlB$_2$
  ($T=\Fe$, Mn, Cr, Co, and Ni) and their alloys}

\author{Liqin Ke}
\email[Corresponding author: ]{liqinke@ameslab.gov}
\author{Bruce N. Harmon}
\author{Matthew J. Kramer}
\affiliation{Ames Laboratory, U.S. DOE, Ames, Iowa 50011, USA}

\begin{abstract}
The electronic structure and intrinsic magnetic properties of
{\fealb}-related compounds and their alloys have been investigated
using density functional theory. For {\fealb}, the crystallographic
$a$ axis is the easiest axis, which agrees with experiments. The
magnetic ground state of {\mnalb} is found to be ferromagnetic in the
basal $ab$ plane, but antiferromagnetic along the $c$ axis. All $3d$
dopings considered decrease the magnetization and Curie temperature in
{\fealb}. Electron doping with Co or Ni has a stronger effect on the
decreasing of Curie temperature in {\fealb} than hole doping with Mn
or Cr. However, a larger amount of Mn doping on {\fealb} promotes the
ferromagnetic to antiferromagnetic transition. A very anisotropic
magnetoelastic effect is found in {\fealb}: the magnetization has a
much stronger dependence on the lattice parameter $c$ than on $a$ or
$b$, which is explained by electronic-structure features near the
Fermi level. Dopings of other elements on B and Al sites are also
discussed.
\end{abstract}

\eid{identifier}
\date{\today}

\maketitle


\section{Introduction}

Magnetic cooling, which is based on the magnetocaloric effect (MCE)
and was discovered one century ago, has long been used in scientific
laboratories to attain extremely low temperatures. A major
breakthrough came in the late 1990s when Pecharsky and Gschneidner
discovered giant MCE around room temperature (RT) in a new class of
magnetic materials~\cite{pecharsky.prl1997}. This discovery has
rekindled research interest in utilizing MCE for much broader
applications, such as domestic appliances, which usually operate
around RT. If successful, this more energy-efficient and
environment-friendly magnetic cooling technique may replace
conventional compressor-based refrigeration and revolutionize the
cooling industry. This new era may arrive in the near
future\textemdash only if one can find or engineer a proper MCE
material which has large MCE under a magnetic field that can be
generated by permanent magnets, and is also abundant, affordable, and
has a good lifespan.

{\fealb} is one of the promising candidates for this purpose and has
attracted great attention since the recent discovery of its
substantial MCE around RT~\cite{tan.jacs2013}. The reported entropy
change has a value of
$\Delta{S_m}=\SIrange{4.1}{7.7}{\J\per\kg\per\K}$ in the presence of
an external field $B=\SIrange{2}{5}{\tesla}$. Although {\fealb} does
not have the largest MCE of all materials, it does not contain any
rare, expensive, or toxic elements. Moreover, its volume barely
changes during the magnetic transition~\cite{lewis.jac2015}, which may
ensure {\fealb} has a good life span for refrigerator applications
operating at high cycle frequencies~\cite{du.jjap2015,du.jpdap2015}.

The MCE often peaks at the Curie temperature $\tc$ of the material;
however, real applications require materials with a large MCE over a
certain operating temperature range. This likely needs to be achieved
by using composite materials with multiple compositions, so the system
can have MCE over the whole operating temperature range for specific
applications. It seems the first logical selection of doping would be
the substitution of Fe with other $3d$ transition-metal elements
$T$. Pure {\mnalb}~\cite{becher.1966} and
{\cralb}~\cite{chaban.im1973} can be formed and share the same
structure of {\fealb}. Combining theory with experiments, K\'adas
$\etal$ studied the phase stability in {\tmalb} with $T=\Cr$, Mn, Fe,
Co, and Ni. They found that although compounds are metastable with
$T=\Co$ and Ni, Fe$_{2-x}$Co$_x$AlB$_2$, Fe$_{2-x}$Ni$_x$AlB$_2$, or
even (Fe$_{2-x-y}$Co$_x$Ni$_y$)AlB$_2$ could be
stable~\cite{kadas.jpcm2017}. However, the magnetic properties of
those alloys or even their parent compounds are not well
understood. For example, the magnetic ground state of {\mnalb} had
been reported to be ferromagnetic (FM)~\cite{becher.1966}, but recent
experiments concluded that it should be either nonmagnetic (NM) or
antiferromagnetic (AFM)~\cite{chai.jssc2015}. To provide guidance on
tuning {\fealb}, a better understanding of the magnetic properties of
pure {\tmalb} and their alloys is desired.

\redtx{Phase stability:} The high melting temperature of FeB makes it
a difficult impurity to remove from {\fealb} samples. The rapid
cooling by melt spinning had been used to greatly improve the {\fealb}
purity by suppressing the growth of FeB~\cite{du.jjap2015}. On the
other hand, extra Al is often added during the synthesis to decrease
the formation of FeB impurities~\cite{du.jjap2015,hirt.ic2016}. With a
higher Al content, Al$_{13}$Fe$_{4}$ becomes the main impurity
phase. Reported magnetization values measured at low temperature vary
between 1.0 and \SI{1.32}{\mu_B\per{Fe}} and the $\tc$ values vary
between 282 and
\SI{320}{\K}~\cite{tan.jacs2013,du.jjap2015,chai.jssc2015,elmassalami.jmmm2011}.
The variation of experimental values may be due to the existence of an
impurity phase in the {\fealb} sample.

Experimentally, substitution of
Mn~\cite{chai.jssc2015,du.jjap2015,du.jpdap2015} or up to
\SI{15}{\percent} of Co~\cite{hirt.ic2016} on Fe sites had been
reported; a large amount of Mn or Co doping makes the structures
unstable~\cite{du.jjap2015} or phase
inhomogeneous~\cite{chai.jssc2015}. Both dopings decrease the
magnetization and $\tc$ in {\fealb}. A spin-glass state had been found
in Fe$_{1.5}$Mn$_{0.5}$AlB$_2$ at low temperature~\cite{du.jpdap2015}.
With Co doping, $\tc$ and the Co content are linearly correlated,
which makes Co content a convenient parameter to tune the material to
have MCE between RT and \SI{200}{K}~\cite{hirt.ic2016}. A few studies
have been reported to investigate the electronic structures and
structural, electronic, and magnetic properties of
{\tmalb}~\cite{elmassalami.jmmm2011,cheng.cms2014,chai.jssc2015,cedervall.jac2016}.

In this work, using density functional theory (DFT), we investigate
the intrinsic magnetic properties including magnetization, exchange
parameters, Curie temperature, and magnetocrystalline anisotropy in
{\tmalb} and their alloys. The magnetoelastic effect in {\fealb} and
the dopings of various elements on B and Al sites are also
discussed. Electronic structures are studied to understand the
magnetic properties.

\section{Computational details}

\begin{figure}[tbp]
\includegraphics[width=1.0\linewidth,clip,angle=0]{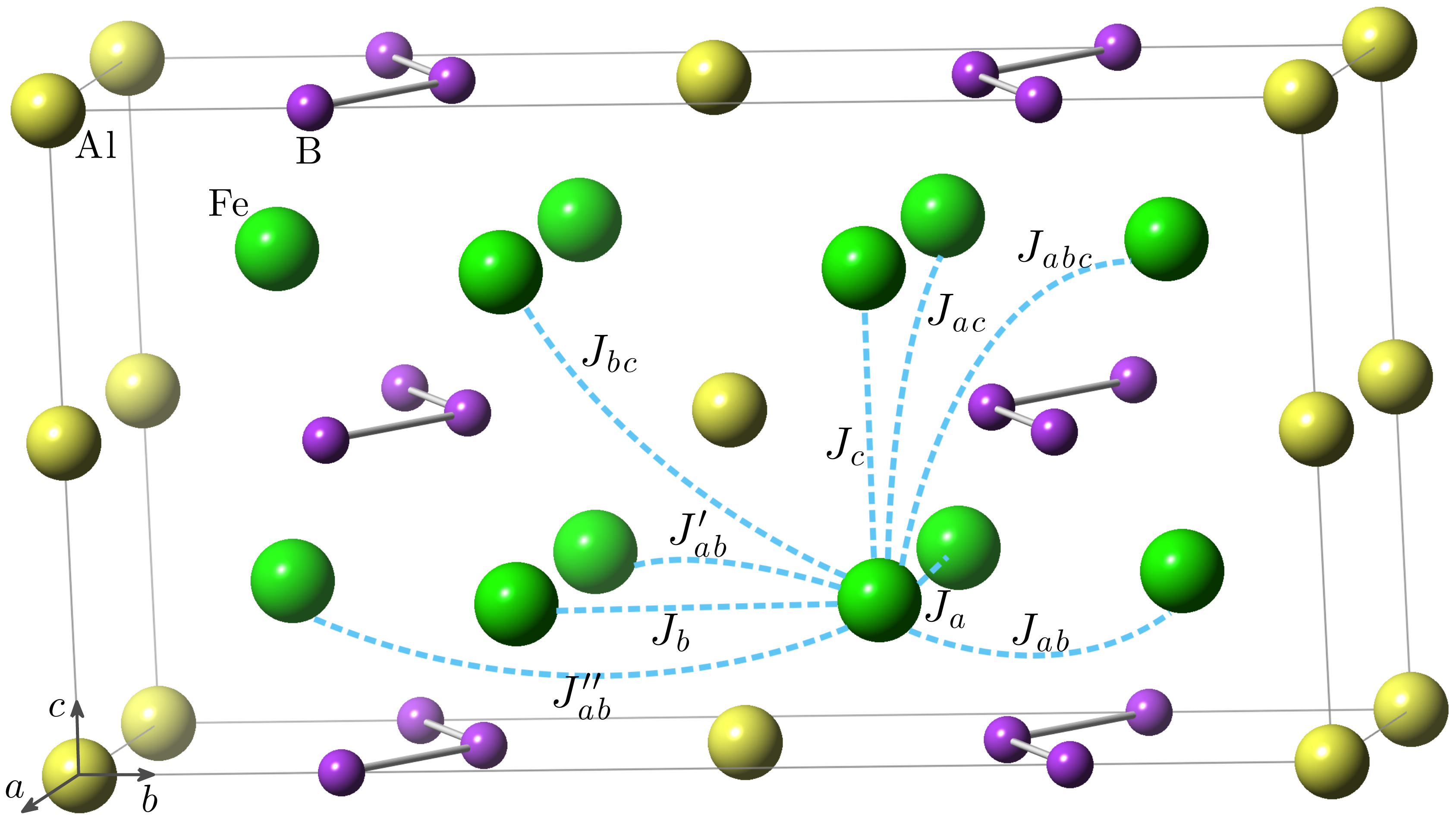}
\caption{Schematic representation of the crystal structure of
  \fealb. The conventional cell is doubled along the $c$ axis to
  depict pair exchange parameters for the first few neighbors of Fe
  atoms (large green spheres). Al atoms are indicated with yellow
  spheres. B atoms, indicated by small purple spheres, form zigzag
  chains along the $a$ axis. Exchange parameters are labeled according
  to the axis or plane of the connecting vector.}
\label{fig:xtal}
\end{figure}

\subsection{Crystal structure}
{\fealb} crystallizes in the orthorhombic \mnalb-type ($Cmmm$, space
group no. 65) structure. The primitive cell contains one formula unit
(f.u.) while the conventional cell contains two.  The crystal
structure is shown in \rfig{fig:xtal}. We double the conventional cell
along the $c$ axis to denote the first few exchange parameters. B
atoms occupy the $4i(m2m)$ site, forming a zigzag chain in the $ab$
plane and along the $a$ axis. Fe atoms occupy the $4j(m2m)$ sites and
form Fe$_6$B prisms with neighboring B atoms. The structure can be
derived from the FeB structure by inserting one layer of Al atoms
perpendicular to the $b$ axis, between each pair of planes containing
the Fe$_6$B prisms~\cite{jeitschko.acta1969}. Each Al atom, which
occupies the $2a(mmm)$ site, is surrounded by eight Fe atoms.
Together they form a body-centered-tetragonal cell elongated along the
$b$ axis. For this Fe-Al cage, the Fe-Al bond length is \SI{2.61}{\AA}
and the Fe-Fe distances are \SIlist{2.87;2.92;3.22}{\AA}, along the
$c$, $a$, and $b$ directions, respectively. Lying in the $ab$ plane,
the nearest Fe-Fe bond has a length of \SI{2.72}{\AA}. The structure
can also be derived by stacking the pure Fe plane and the Al-B plane
alternatively along the $c$ axis. This view is probably more
convenient to understand the magnetic properties, such as exchange
coupling and magnetoelastic effect, both of which are very
anisotropic, behaving very differently along the $c$ axis than along
the $a$ or $b$ axis, as we will discuss later.

\subsection{Computational methods}

Electronic structure and most magnetic properties are calculated using
a standard linear muffin-tin orbital (LMTO) basis
set~\cite{andersen.prb1975} generalized to full potentials
(FP)~\cite{methfessel.chap2000}. This scheme employs generalized
Hankel functions as the envelope functions. Calculations are carried
out within the generalized gradient approximation (GGA) to DFT with
the exchange-correlation parametrization of Perdew, Burke, and
Ernzerhof (PBE)~\cite{perdew.prl1996}, unless
LDA~\cite{vonbarth.jpcssp1972} (local density approximation, with the
exchange-correlation parametrization of von Barth and Hedin) is
specified.

The magnetocrystalline anisotropy energy (MAE) is calculated using the
force theorem~\cite{mackintosh1980}. Starting from the self-consistent
scalar-relativistic potential, the spin-orbit coupling (SOC) is
included in a subsequent one-step calculation with spin being along
direction $\hat{\bf{n}}$. The MAE is characterized below as
$K_{\hat{\bf{n}}}=E_{\hat{\bf{n}}}-E_{001}$, where $E_{001}$ and
$E_{\hat{\bf{n}}}$ are the summation of occupied band energies for the
magnetization being oriented along the $[001]$ and $\hat{\bf{n}}$
directions, respectively.

Exchange coupling parameters $J_{ij}$ are calculated using a static
linear-response approach implemented in a Green's function (GF) LMTO
method, simplified using the atomic sphere approximation (ASA) to the
potential and density~\cite{ke.prb2012,ke.prb2013}. The
scalar-relativistic Hamiltonian is used so SOC is not included,
although it is a small perturbation on $J_{ij}$'s. In the basis set,
$s, p, d$ orbitals are included for $T$ and Al atoms, and $s, p$
orbitals are included for the B atom. A dense $k$-point mesh is used
to calculate exchange parameters $J({\bf q})$, e.g., a $32^3$
$k$-point mesh for the five-atom cell. The real-space $J({\bf R})$ are
obtained by a subsequent Fourier transform. Curie temperatures are
estimated in the mean-field approximation (MFA) with
$k_\text{B}\tc=2/3\sum_i J_{0i}$. The coherent potential approximation
(CPA) implemented within the LMTO-ASA-GF code is used to address the
chemical effects of doping on magnetization and $\tc$. Without using
supercell calculations, the CPA provides an elegant and efficient
approach to investigate substitutional effects with an arbitrary
composition. The details of the methods and applications can be found
elsewhere~\cite{ke.prb2013,ke.prb2016a}.

Both experimental and theoretically optimized crystal structures are
used to investigate the magnetic properties. We fully relax internal
atomic positions and lattice constants with the PBE functional using a
fast plane-wave method, as implemented within the Vienna $\abinitio$
simulation package ({\footnotesize
  VASP})~\cite{kresse.prb1993,kresse.prb1996}. The nuclei and core
electrons are described by the projector augmented-wave
potential~\cite{kresse.prb1999} and the wave functions of valence
electrons are expanded in a plane-wave basis set with a cutoff energy
of up to \SI{520}{eV}.

\section{Results and discussion}

\subsection{Pure compounds: Exchange coupling, magnetic anisotropy, and spin configurations}

\begin{table}[btp]
\caption{Calculated atomic spin magnetic $m_i$, spin magnetization
  $M$, and Curie temperature $\tc$ in {\fealb}. }
\label{tbl:mi}
\bgroup
\def\arraystretch{1.1}
\begin{tabular*}{\linewidth}{l @{\extracolsep{\fill}} crrrrccc}
\hline\hline
\\[-1em]
  {\fealb} &  &   \multicolumn{3}{c}{$m_{\text{i}}$ (\si{\mu_B\per atom})}  &    & $M$ & & $\tc$ \\
\\[-1em]  
  \cline{1-1} \cline{3-5} \cline{7-7} \cline{9-9}
\\[-1.1em]
  Method   &  & Fe   & Al    & B     &  & (\si{\mu_B\per{f.u.}}) &  & (K) \\
\\[-1.1em]
  \hline                                      
\\[-1.1em]
FP-GGA   &  & 1.43 & $-0.04$ & $-0.01$ &  & 2.73 &  &  \\
ASA-GGA  &  & 1.38 & $-0.08$ & $-0.04$ &  & 2.62 &  &  329\\
FP-LDA   &  & 1.31 & $-0.01$ & $-0.03$ &  & 2.54 &  &  \\
ASA-LDA  &  & 1.20 & $-0.06$ & $-0.03$ &  & 2.29 &  &  232\\
\\[-1.0em]
\hline\hline
\end{tabular*}
\egroup
\end{table}

\rtbl{tbl:mi} shows the atomic spin moments $m_i$ at each sublattice
and magnetization $M$ in {\fealb}, which are calculated using the
experimental lattice constants and atomic position parameters from
Ref.~[\onlinecite{cenzual.acta1991}]. Al and B have small moments
antiparallel to the Fe sublattice. Within the GGA, a magnetization of
$M=\SI{1.36}{\mu_B\per Fe}$ is obtained using FP. For the sake of
comparison, we carry out similar calculation for the parent compound,
FeB, and obtain a magnetization of $M=\SI{1.20}{\mu_B\per Fe}$. The
smaller Fe moment in FeB is likely due to its smaller Fe-Fe bond
length (\SI{2.62}{\AA}) than in {\fealb}. ASA gives a slightly smaller
(by \SI{4}{\percent}) magnetization than FP in {\fealb}, suggesting
that ASA is suitable for this material. The calculated
$\tc=\SI{329}{\K}$ is slightly above the upper bound of the reported
experimental $\tc$ values. The agreement is fair considering the MFA
generally overestimates $\tc$. LDA gives smaller magnetization,
especially with ASA, resulting in a smaller $\tc$.

\begin{figure}[tbp]
\begin{tabular}{c}%
  \includegraphics[width=1.0\linewidth,clip,angle=0]{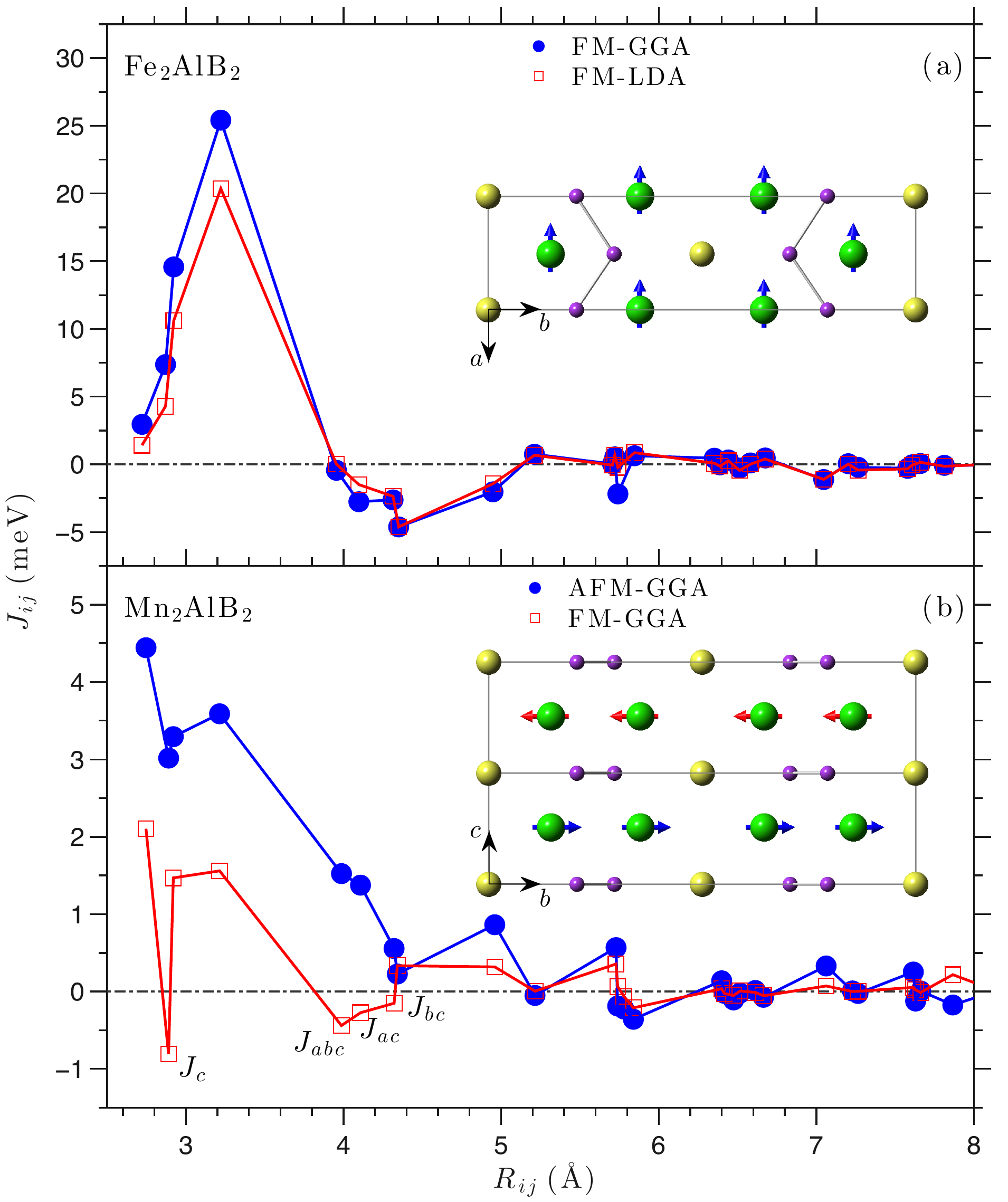}
\end{tabular}
\caption{ Real-space magnetic exchange parameters $\jij$ in {\fealb}
  (a), and {\mnalb} (b) as functions of distance. For {\fealb}, both
  GGA and LDA results are shown. For {\mnalb}, both FM and AFM spin
  configurations are calculated within GGA. The spin configurations of
  their magnetic ground states are shown in the insets.}
\label{fig:jij}
\end{figure}

\begin{table*}[tbp]
\caption{Pairwise exchange parameters $\jij$ for the Heisenberg
  Hamiltonian $H=-\sum J_{ij}\,\hat{\bf e}_i \cdot \hat{\bf e}_j$, and
  $\hat{\bf e}_i$ is the unit vector pointing along the direction of
  the local spin moment at site $i$. The experimental lattice
  parameters and atomic positions are used. For {\fealb}, both GGA and
  LDA results are shown. For {\mnalb}, the PBE functional is used, and
  both FM and AFM spin configurations are considered.}
\label{tbl:jij}%
\bgroup
\def\arraystretch{1.1}
\begin{tabular*}{\textwidth}{l @{\extracolsep{\fill}} lllrrlrrrlrr}
  \hline  \hline
  \\[-1em]
  \multicolumn{3}{c}{Fe$_2$AlB$_2$}  &  & \multicolumn{2}{c}{$R_{ij}$}  &  & \multicolumn{3}{c}{$\hat{\bf R}_{ij}$}   &  & \multicolumn{2}{c}{$J_{ij}$(meV)}\\
  \\[-1em]
  \cline{1-3} \cline{5-6}  \cline{8-10}   \cline{12-13}
  \\[-1.1em]
  Lbl. &  & No. &  & (\AA) & ($a$) &  & $x$ & $y$ & $z$ &  & GGA & LDA\\
  \\[-1.1em]
  \hline
  \\[-1.1em]
  $J_{ab}$   &  & 2 &  & 2.721 & 0.931 &  & 0.5 & $0.785$  & 0     &  & $ 2.95$ & $ 1.41$\\
  $J_c$      &  & 2 &  & 2.870 & 0.982 &  & 0   & $0$      & 0.982 &  & $ 7.36$ & $ 4.29$\\
  $J_a$      &  & 2 &  & 2.923 & 1     &  & 1   & $0$      & 0     &  & $14.58$ & $10.63$\\
  $J_b$      &  & 1 &  & 3.222 & 1.102 &  & 0   & $1.102$  & 0     &  & $25.41$ & $20.37$\\
  $J_{abc}$  &  & 4 &  & 3.955 & 1.353 &  & 0.5 & $-0.785$ & 0.982 &  & $-0.43$ & $ 0.01$\\
  $J_{ac}$   &  & 4 &  & 4.097 & 1.401 &  & 1   & $0$      & 0.982 &  & $-2.76$ & $-1.49$\\
  $J_{bc}$   &  & 2 &  & 4.315 & 1.476 &  & 0   & $1.102$  & 0.982 &  & $-2.63$ & $-2.37$\\
  $J_{ab}'$  &  & 2 &  & 4.350 & 1.488 &  & 1   & $1.102$  & 0     &  & $-4.62$ & $-4.62$\\
  $J_{ab}''$ &  & 2 &  & 4.949 & 1.693 &  & 1.5 & $-0.785$ & 0     &  & $-2.03$ & $-1.43$\\
  \\[-1.1em]
  \hline
  \\[-1.1em]
  \multicolumn{3}{c}{Mn$_2$AlB$_2$} &  & \multicolumn{2}{c}{$R_{ij}$} &  & \multicolumn{3}{c}{$\hat{\bf R}_{ij}$}    &  & \multicolumn{2}{c}{$J_{ij}$(meV)}\\
  \\[-1.1em]
  \cline{1-3} \cline{5-6}  \cline{8-10}   \cline{12-13}
  \\[-1.1em]
  Lbl. &  & No. &  & (\AA) & ($a$) &  & $x$ & $y$ & $z$ &  & FM & AFM\\
  \\[-1.1em]
  \hline
  \\[-1.1em]
  $J_{ab}$   &  & 2 &  & 2.747 & 0.941 &  & $0.5 $ & $-0.797$ & $0     $ &  & $ 2.10$ & 4.44\\
  $J_c$      &  & 2 &  & 2.890 & 0.990 &  & $0   $ & $0     $ & $-0.990$ &  & $-0.81$ & 3.02\\
  $J_a$      &  & 2 &  & 2.920 & 1     &  & $1   $ & $0     $ & $0     $ &  & $ 1.47$ & 3.29\\
  $J_b$      &  & 1 &  & 3.213 & 1.100 &  & $0   $ & $1.100 $ & $0     $ &  & $ 1.56$ & 3.59\\
  $J_{abc}$  &  & 4 &  & 3.987 & 1.365 &  & $-0.5$ & $-0.797$ & $0.990 $ &  & $-0.44$ & 1.52\\
  $J_{ac}$   &  & 4 &  & 4.108 & 1.407 &  & $-1  $ & $0     $ & $0.990 $ &  & $-0.27$ & 1.37\\
  $J_{bc}$   &  & 2 &  & 4.322 & 1.480 &  & $0   $ & $1.100 $ & $0.990 $ &  & $-0.15$ & 0.56\\
  $J_{ab}'$  &  & 2 &  & 4.342 & 1.487 &  & $-1  $ & $1.100 $ & $0     $ &  & $ 0.33$ & 0.23\\
  $J_{ab}''$ &  & 2 &  & 4.960 & 1.698 &  & $-1.5$ & $-0.797$ & $0     $ &  & $ 0.32$ & 0.86\\
  \\[-1.0em]
  \hline\hline
\end{tabular*}
\egroup
\end{table*}

Starting from the FM configuration and using experimental crystal
structures, we calculate the exchange coupling $\jij$ in {\fealb} and
{\mnalb}. Figure \ref{fig:jij} shows the $\jij$ as a function of the
distance $R_{ij}$. In both compounds $\jij$ becomes negligible after
$R_{ij}>\SI{6}{\AA}$. The exchange parameters between the first few
nearest neighbors are also listed in \rtbl{tbl:jij}. Here, $\jij$ can
be treated as stability parameters and a negative $\jij$ indicates
that the given spin configuration is not favorable for that particular
pair of sites~\cite{ke.prb2013}.

For {\fealb}, all of the first four nearest exchange parameters are
positive. The $\jij$ value increases with distance, reaching maximum
at $J_b$, and then decrease, which generally agrees with the previous
study~\cite{cedervall.jac2016}. LDA gives a similar trend but a
smaller amplitude of $\jij$ than GGA, which reflects the smaller
magnetic moments obtained within LDA.

The magnetic ground state of {\mnalb} is not well
understood~\cite{chai.jssc2015}. For simplicity, we start from the FM
configuration. The calculated magnetization is
\SI{0.42}{\mu_B\per{Mn}}, which agrees well with a previous FM
calculation~\cite{chai.jssc2015}. Exchange parameters calculated in
the FM configuration show a very interesting feature: all dominant
$\jij$ are positive for neighbors within the Mn $ab$ plane but
negative for neighbors between neighboring Mn layers, namely, $J_{c}$,
$J_{abc}$, $J_{ac}$, and $J_{bc}$. This suggests that the FM coupling
of Mn atoms is stable within the $ab$ layer but not between
neighboring layers. To confirm it, we calculate $\jij$ for the AFM
configuration, in which FM Mn $ab$ layers couple antiferromagnetically
along the $c$ axis. This AFM configuration gives lower energy than the
FM configuration. Moreover, as shown in \rfig{fig:jij}(b), the
dominant exchange parameters become all positive and larger. To better
quantify the relative stability of those two spin configurations, we
fully relaxed the structure and found that the AFM configuration
increases the on-site Mn moment to \SI{0.75}{\mu_B\per Mn} and lowers
the total energy by \SI{42}{\meV\per{f.u.}} Unlike {\fealb}, the
largest exchange interaction in {\mnalb} is $J_{ab}$, which is between
the nearest Mn neighbors. The N\'eel temperature is estimated to be
$\tn=\SI{310}{K}$ within MFA for the AFM configuration with the
experimental crystal structure.

\begin{figure}[tbp]
\begin{tabular}{c}%
  \includegraphics[width=1.0\linewidth,clip,angle=0]{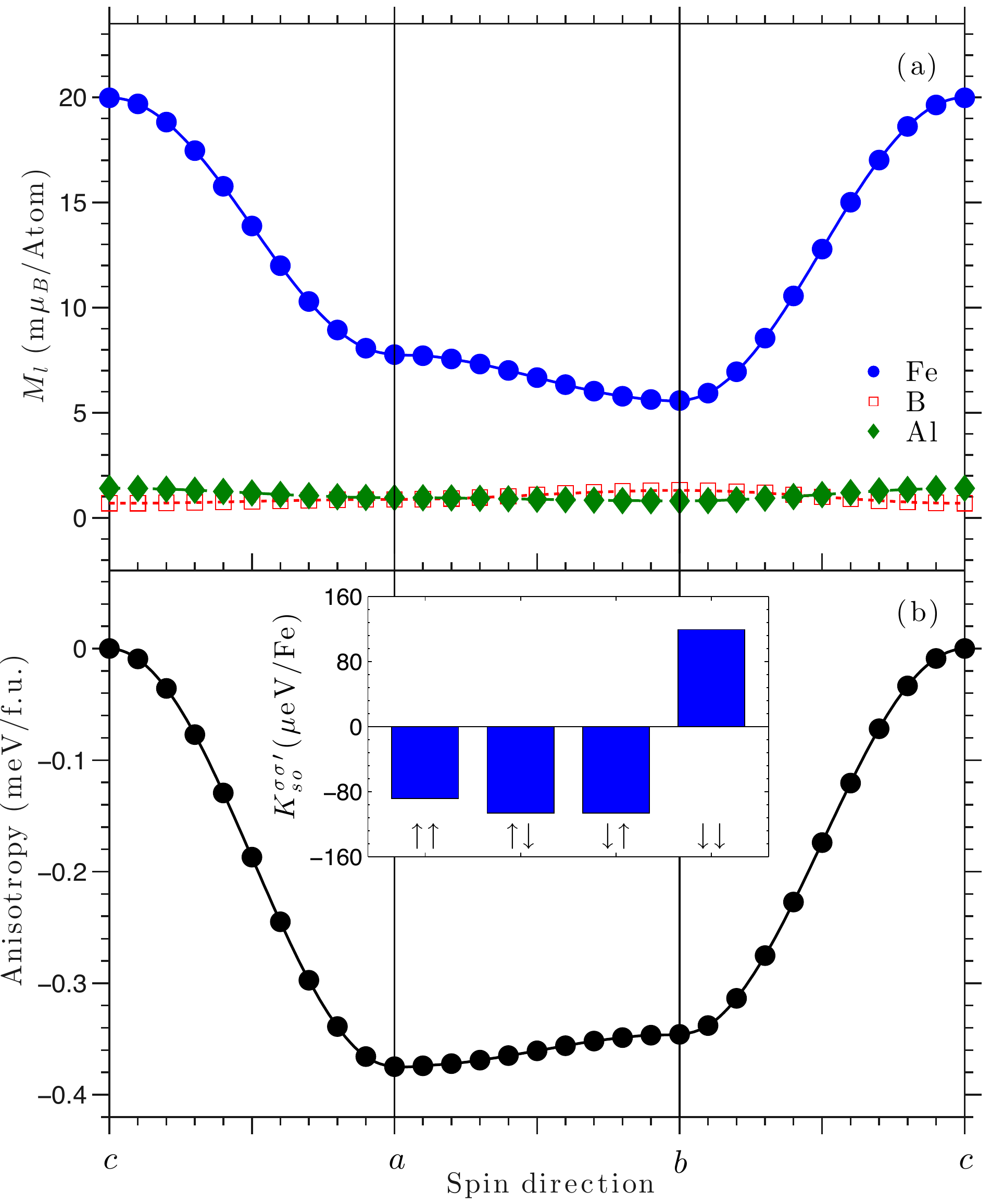}
\end{tabular}
\caption{ Variation of (a) atomic orbital magnetic moments of Fe, Al,
  and B sublattices, and (b) energy as functions of spin quantization
  axis rotation in {\fealb}. The inset in panel (b) shows the
  spin-resolved anisotropy of Fe-site spin-orbit coupling energy
  $K_\text{so}^{\sigma\sigma'}=\frac{1}{2}\langle V_\text{so}
  \rangle_{110}{-}\frac{1}{2}\langle V_\text{so}\rangle_{001}$.}
\label{fig:k_kl}
\end{figure}

Next we consider the SOC effect in {\fealb}. \frfig{fig:k_kl} shows
the energy and the average orbital magnetic moments of each sublattice
as functions of spin quantization axis orientation, which rotates from
the $c\to a\to b\to c$ axis. The relaxed structure was used for the
calculations. The $a$ axis is the easiest axis, which agrees with
recent neutron scattering experiments~\cite{cedervall.jac2016}. The
$c$ axis is the hardest axis, while the anisotropy within the $ab$
plane is very small. Energy changes by
$K_{100}=\SI{-0.38}{\meV\per{f.u.}}$ (\SI{-1.34}{MJ\per m^3}) when the
spin quantization axis rotates from the $c$ axis to the $a$ axis. The
anisotropy calculated using experimental
structure~\cite{cenzual.acta1991} is larger by
$\sim$\SI{10}{\percent}, reaching $K_{100}=\SI{-1.47}{MJ\per m^3}$. B
and Al atoms have negligible orbital magnetic moments, as expected for
light $2p$ and $3p$ elements. The orbital moment of Fe is also rather
small, and interestingly has the largest value of
\SI{0.02}{\mu_B\per{Fe}} when spin is along the hardest $c$ axis
instead of in the $ab$ plane. Similar behavior has also been found in
materials such as FePt~\cite{antropov.ssc2014}. To elucidate the
origin of MAE, we evaluate the anisotropy of the on-site SOC energy
with respect to the $c$ axis and the $ab$ plane,
$K_{110}^\text{so}=\frac{1}{2}\langle
V_\text{so}\rangle_{110}{-}\frac{1}{2}\langle
V_\text{so}\rangle_{001}$, and resolve it into four spin
channels~\cite{ke.prb2015,ke.prb2016b}. It is well known that only
when the MAE is dominated by the $\dd$ term, one may expect an obvious
correlation between the orbital moment and MAE, and a larger orbital
moment along the easy axis~\cite{bruno.prb1989,ke.prb2015}. As shown
in the inset of \rfig{fig:k_kl}(b), the amplitudes of four spin
components of $K_\text{so}^{\sigma\sigma'}$ are comparable. The $\dd$
term favors the spin to be along the $c$ axis, while the other three
terms ($\uu$, $\ud$, and $\du$) favor the spin to lie in the $ab$
plane, explaining the absence of the correlation between MAE and
orbital moment in {\fealb}. LDA gives a smaller anisotropy
($K_{100}=\SI{-0.82}{MJ\per m^3}$ using experimental structure) but a
similar trend of angular dependence of energy. For {\mnalb}, the $c$
axis is also the hardest axis. Within the $ab$ plane, the system has
slightly lower energy when the spin is along the $b$ axis. As
expected, its anisotropy is much smaller than in {\fealb}. The
schematic representation of the ground-state spin configurations of
{\fealb} and {\mnalb} are shown in the insets of \rfig{fig:jij}.

\begin{figure}[tbp]  
\begin{tabular}{c}%
  \includegraphics[width=1.0\linewidth,clip,angle=0]{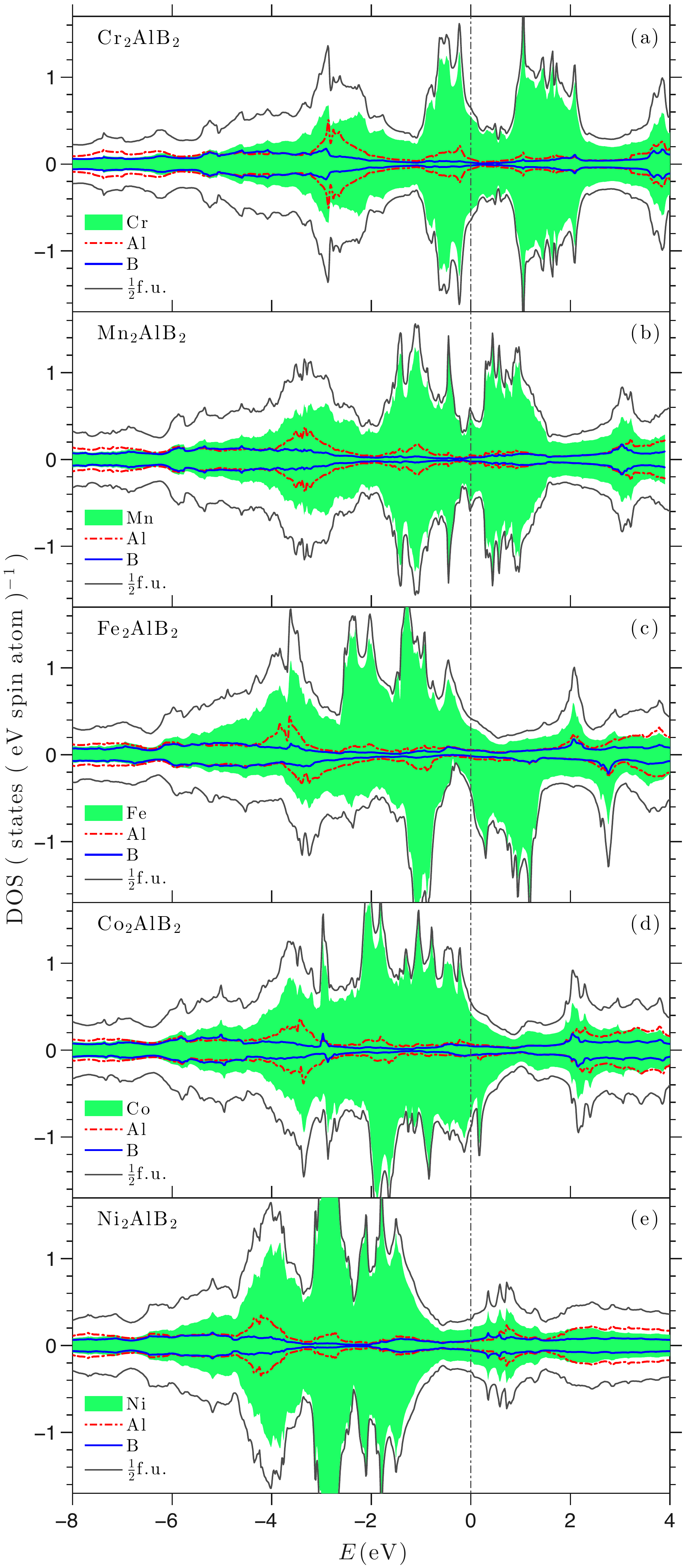} 
\end{tabular}
\caption{ Atom- and spin-projected, scalar-relativistic partial
  densities of states (DOS) in (a) {\cralb}, (b) {\mnalb}, (c)
  {\fealb}, (d) {\coalb}, and (e) {\nialb} calculated within GGA and
  using fully relaxed structures. The total DOS of the f.u. cell are
  scaled by $\frac{1}{2}$ to better compare. {\cralb} and {\nialb} are
  nonmagnetic. {\mnalb} is antiferromagnetic and {\fealb} is
  ferromagnetic. {\coalb} is weakly ferrogmagnetic.  Fermi energy
  $E_F$ is at \SI{0}{\eV}. }
\label{fig:pdos}
\end{figure}

\frfig{fig:pdos} shows the scalar-relativistic partial density of
states (PDOS) projected on individual elements in {\tmalb} with
$T=\Cr$, Mn, Fe, Co, and Ni. The total density of states (DOS), scaled
by $1/2$, is also shown to compare. The total DOS of {\fealb} compares
well with previously reported calculations~\cite{chai.jssc2015}.
Al-$3s$ and B-$2s$ states are located between $-12$ and \SI{-7}{\eV}
below the Fermi level $\efermi$.  Al-$3p$ states hybridize with $T$
atoms at around \SI{-4}{\eV} below $\efermi$. For {\mnalb}, the Fermi
level is located at a pseudogap in the AFM configuration, and the DOS
at $\efermi$ is smaller than in the FM configuration, which again
suggests that the AFM configuration is more stable. The calculated
hypothetical {\coalb} shows a weak magnetic moment of about
\SI{0.2}{\mu_B\per{Co}}. The two peaks at \SI{\pm0.1}{\eV} around the
Fermi level, as shown in \rfig{fig:pdos}(d), will be pinched at
$\efermi$ in the non-spin-polarized calculation (not shown). Thus, the
small spin polarization decreases the DOS at $\efermi$ and stabilizes
the system. The calculated {\cralb} and hypothetical {\nialb} are
nonmagnetic and have small DOS at $\efermi$. It is worth noting that
Ni$_x$B$_{1-x}$ systems become magnetic only after
$x>0.75$~\cite{vandergeest.calphad2014}.

\begin{table}[tbp]
\caption{Lattice parameters, internal atomic positions, $y_{4j}$ and
  $y_{4i}$, on-site atomic magnetic moment of $T$ atoms,
  $m_T$ (\si{\mu_B}), relative total energy (\si{\meV\per{f.u.}}), and
  critical temperatures (Curie temperature in {\fealb} or N\'eel
  temperature in {\mnalb}), $\tc$ (\si{\K}) in $T_2$AlB$_2$ with
  $T=\Fe$, Mn, Cr, Co, and Ni. $T$ atom occupies the $4j$ site (0
  $y_{4j}$ 1/2) and B atom occupies the $4i$ (0 $y_{4i}$ 0) site.}
\label{tbl:lattice_constant}%
\bgroup
\def\arraystretch{1.1}
\begin{tabular}{llllllccc}  
\hline
\hline
\\[-1em]
$T$ & $a$ & $b$ & $c$ & $y_{4j}$ & $y_{4i}$ & $m_T$ & $\Delta E$ & $\tc$ \\
\\[-1em]
\hline
\\[-1.1em]
Cr & 2.921 & 11.034 & 2.929 & 0.3521 & 0.2057 & 0 &  & \\
Expt.~[\onlinecite{chaban.im1973}] & 2.937 & 11.07  & 2.971 & 0.352 & 0.220 &  & &\\
Expt.~[\onlinecite{lu.mrl2016}]    & 2.937 & 11.047 & 2.968 &  &  &  & &\\
\\[-1.1em]
\hline
\\[-1.1em]
Mn-NM & 2.890 & 11.050 & 2.817 & 0.3562 & 0.2060 & 0 & 0 &\\
Mn-FM & 2.892 & 11.056 & 2.826 & 0.3551 & 0.2060 & 0.42 & $-21.5$ &\\
Mn-AFM & 2.887 & 11.109 & 2.830 & 0.3547 & 0.2061 & 0.75 & $-63.6$ & 296\\
Expt.~[\onlinecite{cenzual.acta1991}] & 2.92 & 11.08 & 2.89 & 0.355 & 0.209 &  & &\\
Expt.~[\onlinecite{chai.jssc2015}]  & 2.936 &  11.12 & 2.912 &  &  &  & & \\
\\[-1.1em]
\hline
\\[-1.1em]
Fe-NM & 2.951 & 11.261 & 2.698 & 0.3531 & 0.2065 & 0 & 0 &\\
Fe-AFM & 2.941 & 11.212 & 2.739 & 0.3559 & 0.2070 & 1.06 & $-97.9$ & \\
Fe-FM & 2.915 & 11.017 & 2.851 & 0.3537 & 0.2063 & 1.37 & $-164.1$ & 298 \\
Expt.~[\onlinecite{jeitschko.acta1969}] & 2.923 & 11.034 & 2.870 & 0.3540 & 0.2071 &  & & \\
\\[-1.1em]
\hline
\\[-1.1em]
Co & 2.962 & 11.314 & 2.689 & 0.3541 & 0.2073 & 0.21 & & \\
\\[-1.1em]
\hline
\\[-1.1em]
Ni & 2.979 & 11.041 & 2.843 & 0.3586 & 0.2101 & 0 & & \\
\\[-1em]
\hline
\hline
\end{tabular}
\egroup
\end{table}

Table \ref{tbl:lattice_constant} summarizes the lattice parameters,
atomic positions, atomic moment of $T$ site, relative total energies,
and critical temperatures of {\tmalb} with different magnetic
configurations. The calculated lattice parameters and atomic positions
agree well with experiments. The relaxed lattice parameters of
{\tmalb} not only vary with element $T$ but also depend on the spin
configuration. For {\fealb}, lattice parameters $a$ and $b$ decrease,
and $c$ increases when the spin configuration changes from
$\NM\to\AFM\to\FM$. The calculated lattice parameters using the FM
configuration agree the best with experiments. For {\mnalb}, $b$ and
$c$ increase when the spin configuration changes from
$\NM\to\FM\to\AFM$. Relative to those of {\fealb}, the lattice
parameter $a$ of {\tmalb} varies within \SI{0.06}{\AA}
(\SI{2.2}{\percent}) in the sequence of $\Mn<\Fe\approx\Cr<\Co<\Ni$;
$b$ varies within \SI{0.30}{\AA} (\SI{2.7}{\percent}) in the sequence
of $\Fe<\Ni<\Cr\approx\Mn<\Co$; and parameter $c$ varies within
\SI{0.16}{\AA} (\SI{5.7}{\percent}) in the sequence of
$\Co<\Mn<\Ni\approx\Fe<\Cr$.  Percentagewise, the largest variation
occurs with lattice parameter $c$. As we will show later, {\fealb} has
a much stronger magnetoelastic effect along the $c$ axis.

\subsection{Alloys: $M$ and $\tc$ in {\fetaalb}} 

\begin{figure}[tbp]  
\begin{tabular}{c}%
  \includegraphics[width=1.0\linewidth,clip,angle=0]{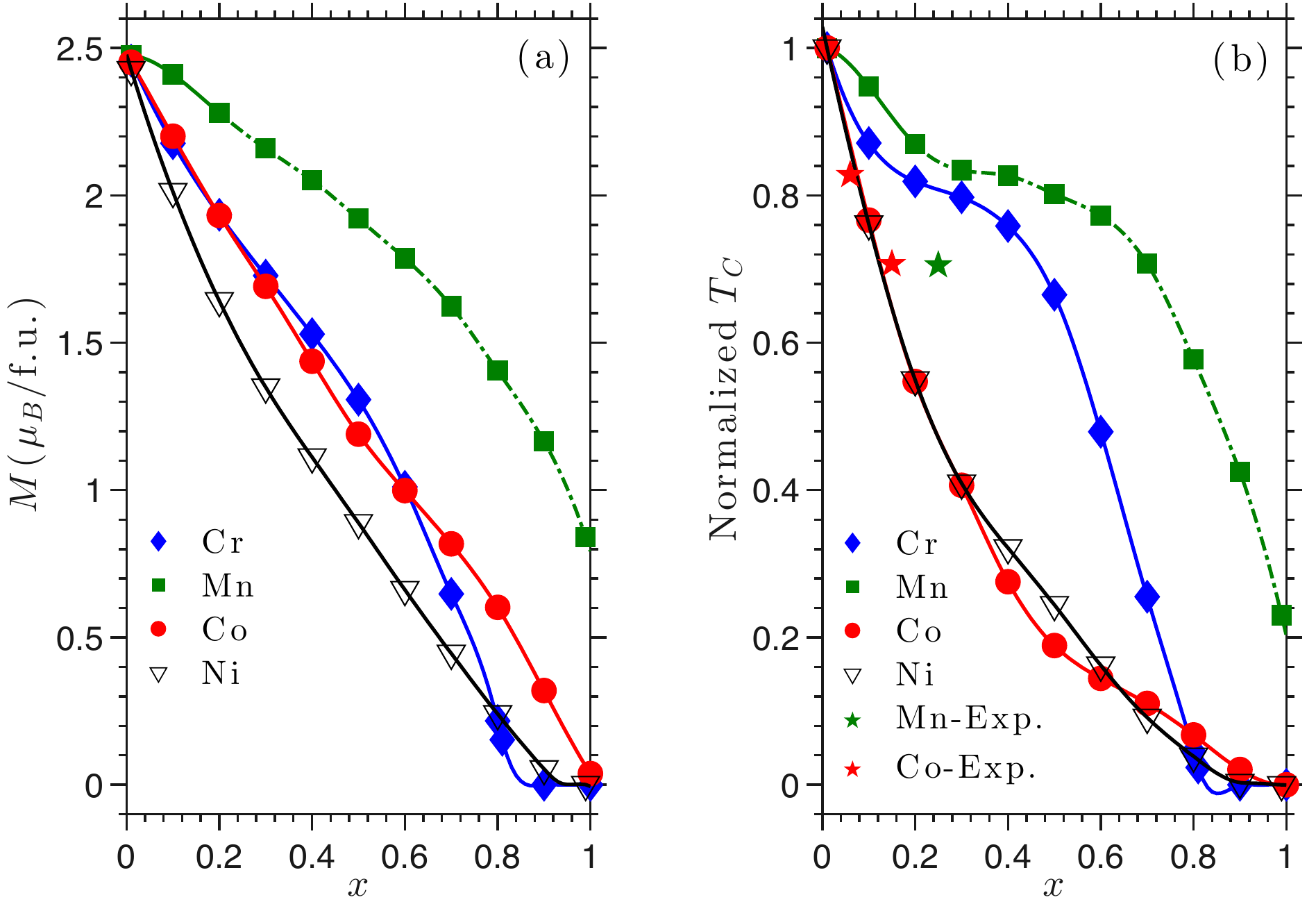}
\end{tabular}
\caption{ Spin magnetization $\mscell$ (a) and normalized $\tc$ (b) as
  functions of doping concentration $x$ in {\fetbalb} with $T=\Cr$,
  Mn, Co, and Ni. Theoretical $\tc$ is estimated within MFA and
  normalized with respect to pure {\fealb}. Experimental $\tc$ values
  of Mn and Co, adopted from Refs.[\onlinecite{du.jpdap2015}] and
  [\onlinecite{hirt.ic2016}], are denoted by stars in green and red,
  respectively. For {\femnalb}, the FM configuration becomes less
  stable than AFM within CPA after $x>0.2$, which are denoted by the
  green dashed line. }
\label{fig:m_j0_vs_x_cpa}
\end{figure}

\begin{figure}[tbp]  
\begin{tabular}{c}%
  \includegraphics[width=1.0\linewidth,clip,angle=0]{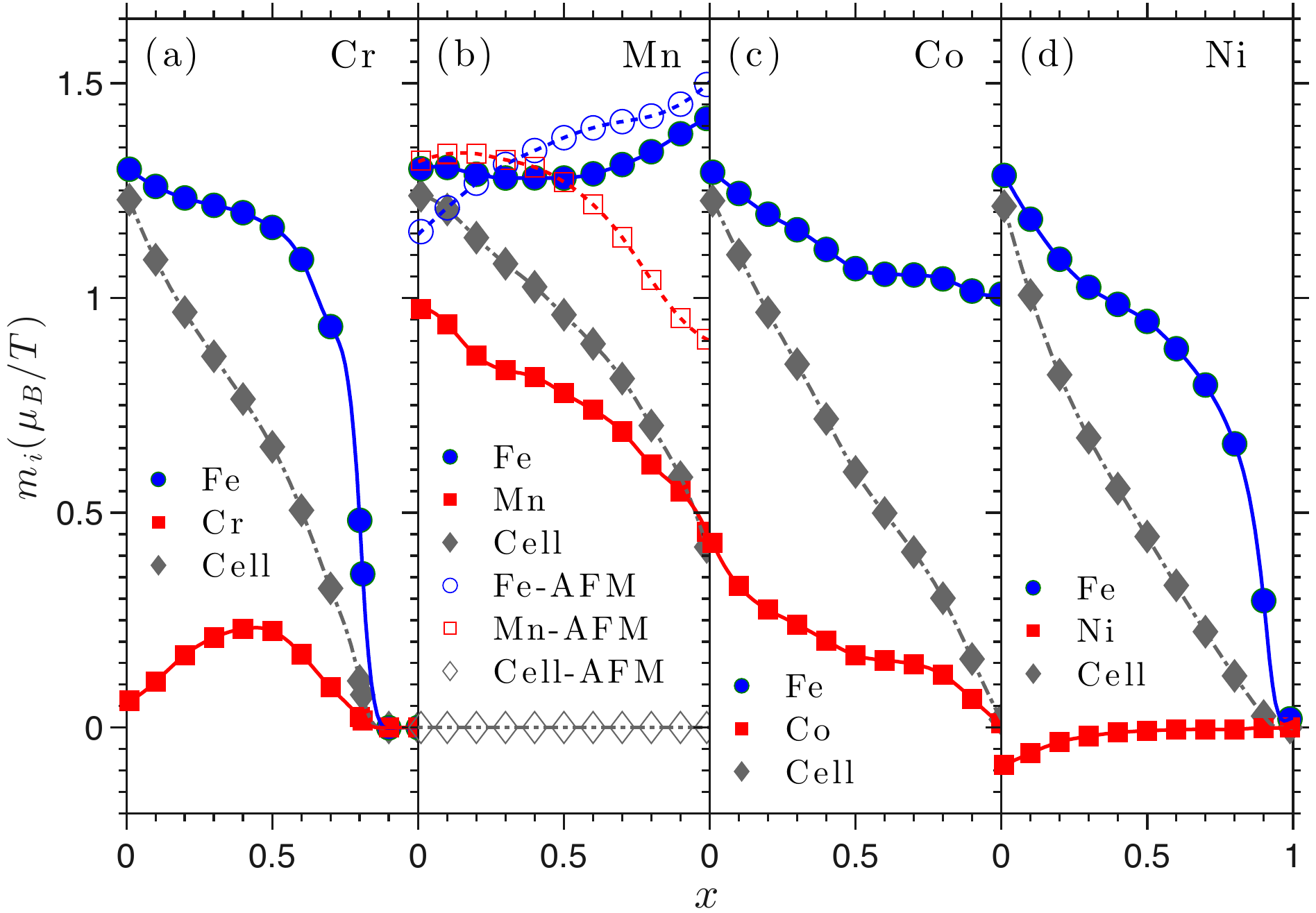}
\end{tabular}
\caption{Component-resolved atomic spin moments $\mspin$ as functions
  of doping content $x$ in {\fetaalb} with $T=\Cr$ (a), Mn (b), Co
  (c), and Ni (d). Calculations are carried out using the LMTO-ASA-CPA
  method. Structural changes due to substitution are neglected, and
  the experimental lattice parameters and atomic positions of {\fealb}
  are used for all calculations. For $T=\Mn$, the AFM configuration is
  also considered besides the FM configuration, and the absolute
  magnetic moments of Mn and Fe components are shown.}
\label{fig:msite_vs_x_cpa}
\end{figure}

To investigate how magnetic properties change with the $3d$
substitutions on Fe sites in {\fealb}, we first consider the chemical
effect by neglecting the structure changes caused by substitution. The
LMTO-ASA-CPA method is used to calculate the magnetization and the
normalized effective exchange (or MFA estimation of $\tc$ in units of
pure {\fealb}) in {\fetaalb} as functions of doping concentration $x$,
with $T=\Cr$, Mn, Co, and Ni. The experimental lattice parameters and
atomic positions of {\fealb} are used and the results are shown in
\rfig{fig:m_j0_vs_x_cpa}.  All dopings decrease the magnetization and
$\tc$ in {\fealb}. The component-resolved atomic spin moments in those
alloys are shown in \rfig{fig:msite_vs_x_cpa}. For Mn doping, we also
consider the AFM configuration and show the absolute values of
component-resolved moments in \rfig{fig:msite_vs_x_cpa}(b).

Assuming the FM configuration, Mn has the slightest effect on the
decrease of the magnetization and $\tc$. The Fe moment barely changes
and even increases with a higher Mn content. The decrease of total
magnetization is due to the dilution of Fe moments with smaller Mn
moments. With a \SI{25}{\percent} Mn content, the calculated $\tc$
decreases by \SI{20}{\percent} while experiments found a larger
$\Delta\tc=\SI{-30}{\percent}$. The assumption of FM configuration is
logical when the Mn amount is small. However, at a higher Mn content,
the AFM configuration should also be considered, given that pure
{\mnalb} is more stable with the AFM configuration. Here we calculate
the AFM configuration in CPA, by assuming the spin moments of $3d$
atoms (both Fe and Mn components) are parallel within the $ab$ plane
and antiparallel between neighboring planes. As shown in
\rfig{fig:msite_vs_x_cpa}(b), in comparison with the FM configuration,
the AFM configuration gives larger Mn moments in the whole doping
range and larger Fe moments at $x\ge0.3$. Within CPA and without
considering any lattice relaxation, the AFM configuration becomes more
stable than the FM configuration with $x>0.2$. Thus, the larger
decrease of $\tc$ observed in experiments is likely caused by the
forming of AFM phases in the samples. By systematically investigating
solid solutions {\femnalb}, Chai $\etal$ observed both NM and FM
M\"ossbauer spectral components in all Mn-containing samples and
attributed them to the clustering of Mn-rich and Fe-rich regions in
the samples. Moreover, a spin-glass state has been observed at low
temperature in {\femnalb} with $x=0.25$ and this phenomenon had been
interpreted as the result of geometric frustration caused by the
triangular configuration of magnetic atoms~\cite{du.jpdap2015}. Here,
we argue that it could be caused by the competition between the FM and
AFM configurations along the $c$ axis.

Given that a large DOS lies right above the Fermi level in the
minority spin channel as shown in \rfig{fig:pdos}(c), it is not
surprising that the electron doping, such as Co or Ni doping,
decreases the magnetic moment on Fe sites. With a small amount of Co
doping, the magnetization and $\tc$ decrease nearly linearly with Co
content. Similar linear dependence of $\tc$ on Co content had been
observed in experiments. As shown in \rfig{fig:m_j0_vs_x_cpa}(b),
calculated {$\Delta\tc$} values agree very well with
experiments~\cite{hirt.ic2016}. With $0.4<x<0.9$, Co atoms in
{\fecoalb} have a small moment of \SI{\sim0.2}{\mu_B\per Co}, similar
to the Co moment calculated in the fully relaxed structure of {\coalb}
using FP. However, this small moment becomes unstable in ASA at $x=1$.

In comparison to Co doping, Ni doping has a similar effect on
decreasing $\tc$ and an even stronger effect on suppressing the
magnetization in {\fealb}. As shown in \rfig{fig:msite_vs_x_cpa}(d),
the atomic Ni moment in {\fenialb} is small and coupled antiparallel
with the Fe sublattice at small $x$, and negligible for $x>0.3$.

In {\fecralb }, the Cr moment is small and parallel to the Fe
sublattice. The maximum Cr moment of \SI{0.2}{\mu_B\per Cr} occurs at
$x=0.5$. Like Mn, Cr doping has a smaller effect on decreasing the
$\tc$ than Co and Ni. However, Cr doping is not likely to promote the
$\FM\to\AFM$ transition, which may compromise the MCE as in the case
of Mn doping~\cite{du.jpdap2015}. Thus, it is worthwhile to
investigate Cr doping, which may provide a useful approach to tune the
$\tc$ and MCE in {\fealb}.

\subsection{Effect of lattice distortion}

\begin{figure}[tbp]  
\begin{tabular}{c}%
  \includegraphics[width=1.0\linewidth,clip,angle=0]{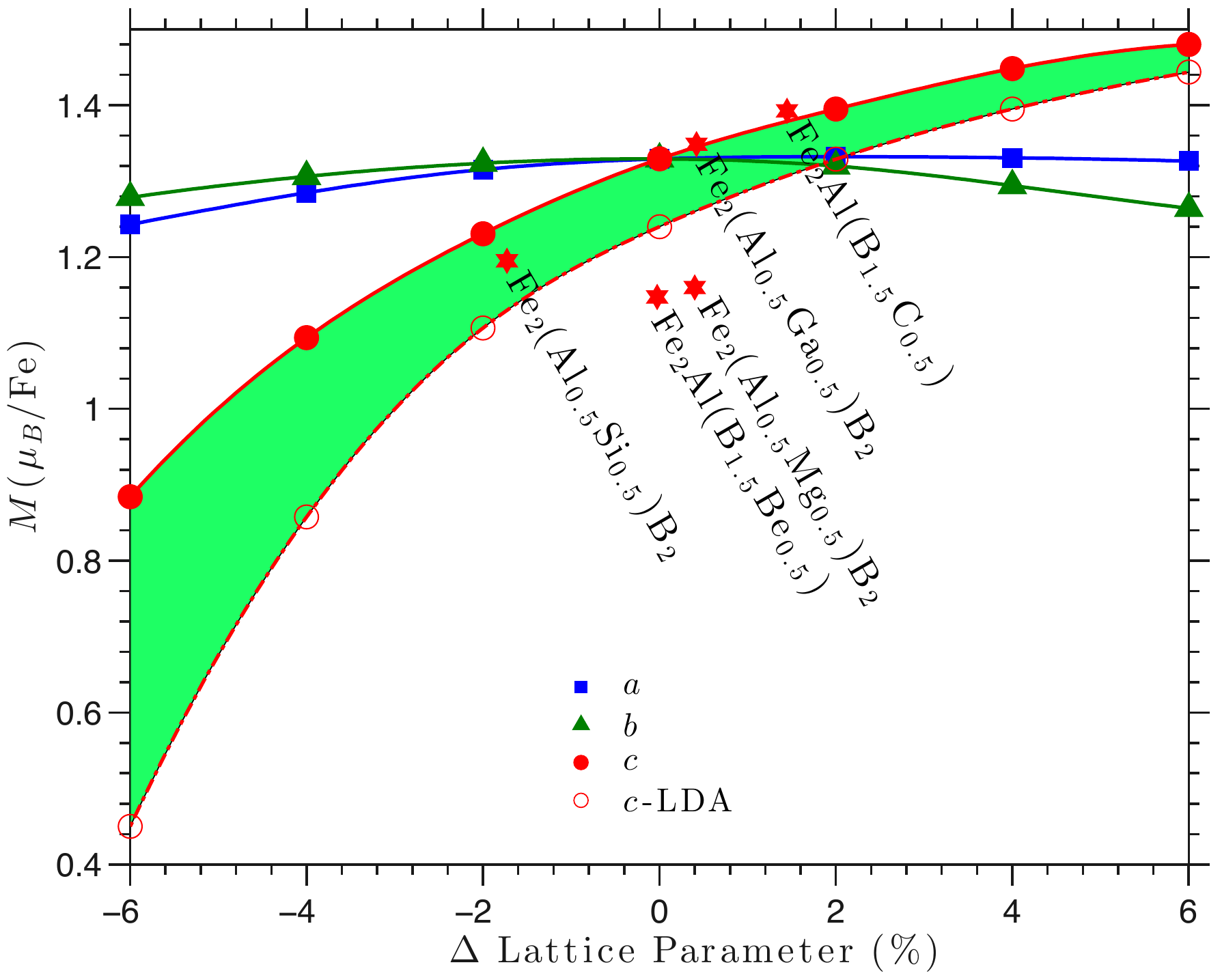}
\end{tabular}
\caption{Magnetization as functions of lattice parameters. Each of the
  three lattice parameters is varied with the other two being
  preserved. For the lattice distortion along the $c$ axis, LDA
  results are also shown to compare. Magnetizations calculated in
  fully relaxed Fe$_2$Al(B$_{1.5}$Be$_{0.5}$),
  Fe$_2$Al(B$_{1.5}$C$_{0.5}$), Fe$_2$(Al$_{0.5}$Mg$_{0.5}$)B$_2$,
  Fe$_2$(Al$_{0.5}$Si$_{0.5}$)B$_2$, and
  Fe$_2$(Al$_{0.5}$Ga$_{0.5}$)B$_2$ are also shown.}
\label{fig:m_vs_strain}
\end{figure}

Besides the chemical effect, the volume change caused by substitution
may also affect the magnetic properties. As shown in
\rtbl{tbl:lattice_constant}, the lattice parameters in {\tmalb} vary
with the element $T$ and spin configuration. To have a rough idea on
the magnetoelastic effect in {\fealb}, we calculate the magnetization
dependence on the three lattice parameters, respectively. Starting
from the fully relaxed structure, each of the three parameters is
varied while the other two are kept constant. Interestingly, as shown
in \rfig{fig:m_vs_strain}, the magnetoelastic effect in {\fealb} is
very anisotropic. The magnetization has a much stronger dependence on
the lattice parameter $c$ than on $a$ or $b$. With $\Delta
c=\SI{-6}{\percent}$, magnetization decreases by \SI{35}{\percent}
within GGA and \SI{60}{\percent} within LDA.

\begin{figure*}[btp]
\begin{tabular}{c}%
\includegraphics[width=0.93\linewidth,clip,angle=0]{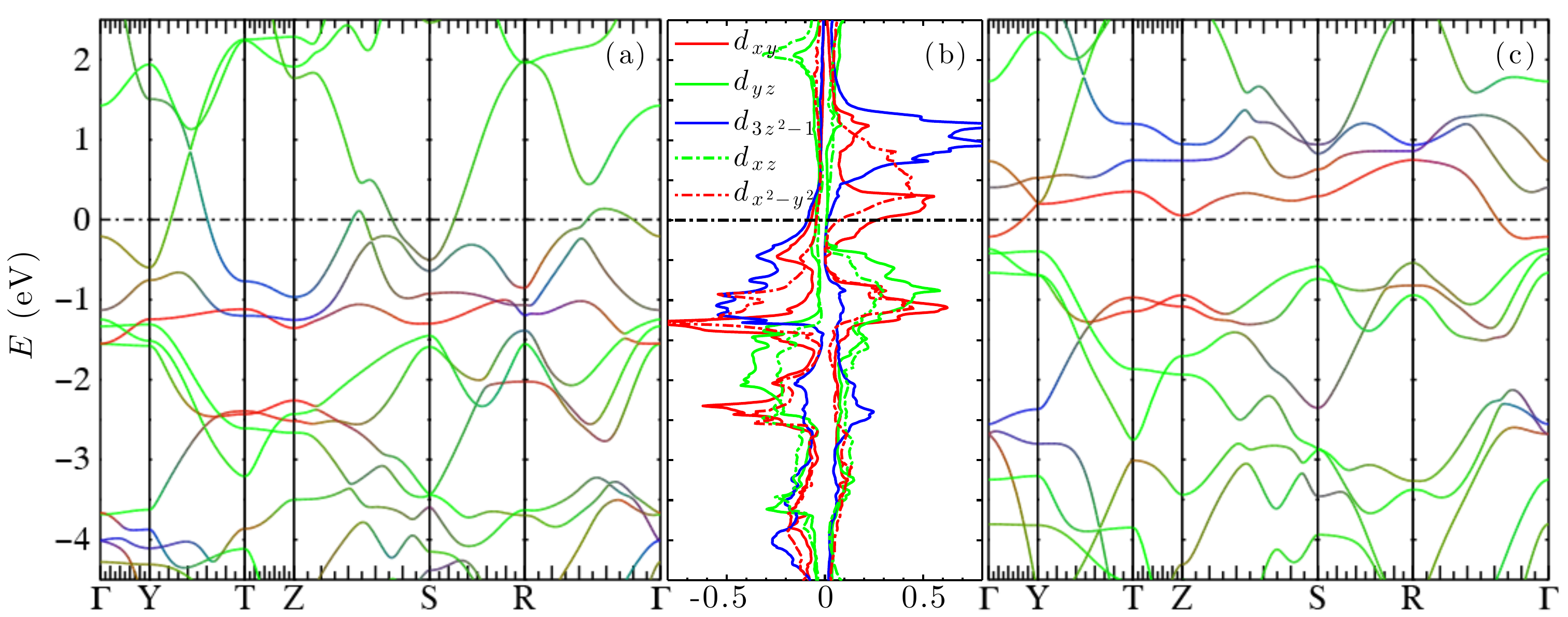} \\
\end{tabular}
\caption{ Band structure of {\fealb} in (a) marjory and (c) minority
  spin channels. Bands are with color weights, with blue identifying
  the Fe-$d_{3z^2-1}$ states, red the Fe-($d_{xy}$, $d_{x^2-y^2}$)
  states, and green everything else. PDOS $\big[$states ( eV~spin~atom
    )$^{-1}$$\big]$ projected on Fe-$3d$ states in two spin channels
  are shown in panel (b). The left and right portions of panel (b)
  show PDOS in the majority and minority spin channels,
  respectively. The horizontal dashed lines in all three panels
  indicate the Fermi level.}
\label{fig:band}
\end{figure*}

\begin{figure}[tbp]  
\begin{tabular}{c}%
  \includegraphics[width=0.99\linewidth,clip,angle=0]{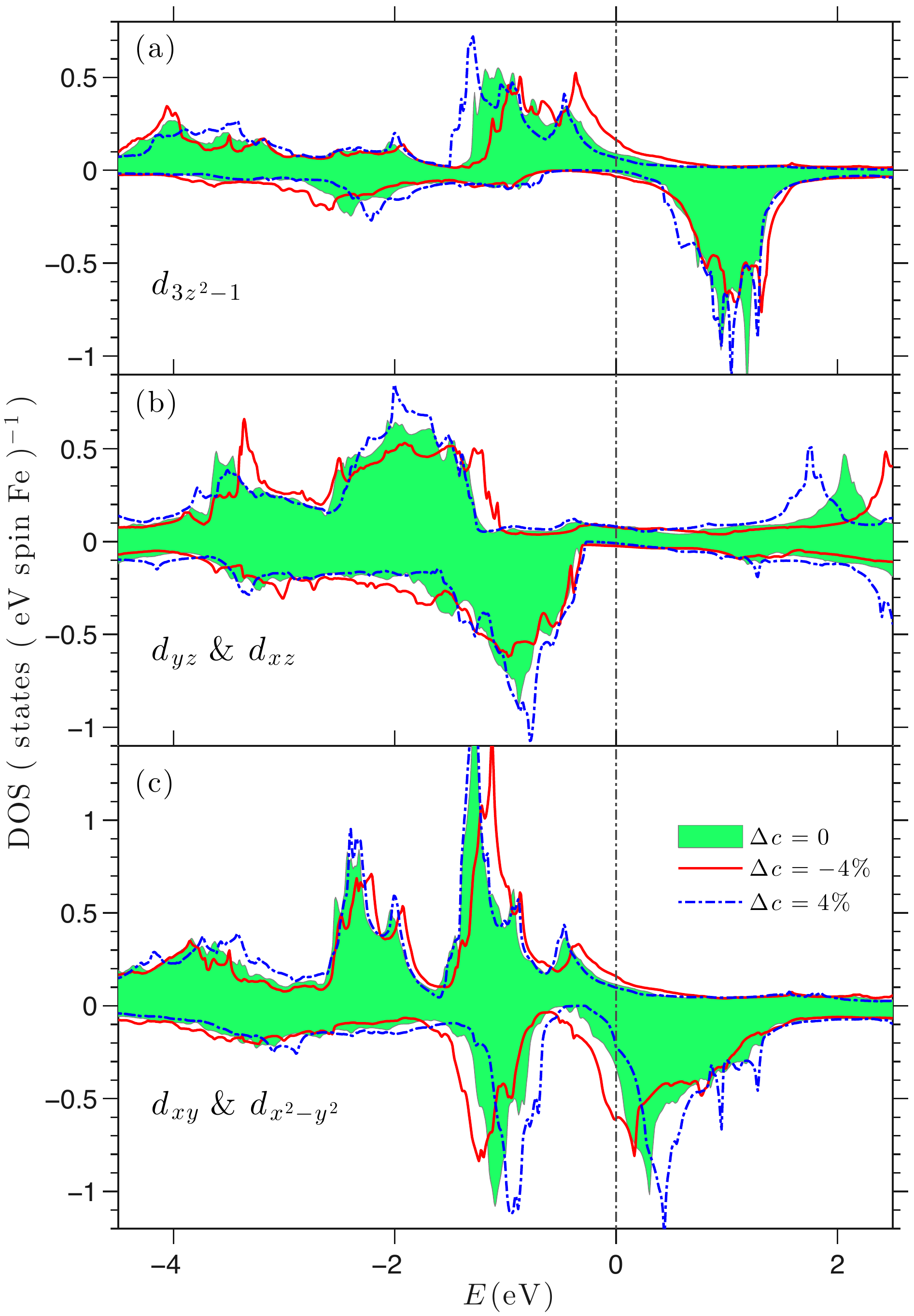}
\end{tabular}
\caption{The partial density of states projected on Fe $3d$ with
  different $c$ parameters.  The scalar-relativistic partial density
  of states projected on the 3$d$ states of Fe sites in {\fealb} with
  different lattice parameters. }
\label{fig:pdos_vs_strain}
\end{figure}

This anisotropic magnetoelastic effect can be understood by
investigating the electronic structure features near $\efermi$ and
their changes caused by the lattice distortion. The band structures of
{\fealb} in the majority and minority channels are shown in
\rfig{fig:band}(a, c), respectively. The PDOS projected on Fe-$3d$
states and in two spin channels are shown in
\rfig{fig:band}(b). Particularly of note is a narrow band right above
$\efermi$ in the minority spin channel. This band is found to consist
almost entirely of Fe-$d_{xy}$ and $d_{x^2-y^2}$ ($m=\pm2$)
orbitals. The variations of PDOS with lattice parameter $c$ are shown
in \rfig{fig:pdos_vs_strain}. The decreasing of parameter $c$
increases the bandwidth of the $d_{3z^2-1}$ state, which has a
relatively large density right below $\efermi$ in the majority spin
channel. These antibonding $d_{3z^2-1}$ states, located between
\SI{-1.3}{\eV} and $\efermi$, shift up toward $\efermi$ and become
less occupied. Correspondingly, the aforementioned peak of $m=\pm2$
states, located right above $\efermi$ in the minority spin channel,
become more occupied. As a result, the magnetization decreases. The
Fe-$d_{yz}$ and $d_{xz}$ states ($m=\pm1$) have small DOS around
$\efermi$ and contribute less to this magnetization change. With the
further decrease of $c$ and then magnetization, the spin splitting
becomes smaller, which quickly accelerates the decrease of
magnetization as $\Delta c$ approaches to \SI{-6}{\percent}.

\subsection{Dopings on B and Al sites}
We also consider the substitutions of B and Al atoms with their
neighboring elements in the periodic table: Be and C atoms on the B
site, and Mg, Si, and Ga atoms on the Al site. The stabilities of
those dopings are not well understood, and a careful and complete
future investigation is desired. Here we focus on the possible effects
of those dopings on the magnetization. Using various configurations of
a ten-atom {\fealb} unit cell, we substitute one B or Al atom with a
dopant atom and fully relax the structures for {\fealbz} with
$Z=\text{Be}$ and C, and {\fealzb} with $Z=\text{Mg}$, Si, and
Ga. Their magnetizations calculated with the corresponding lowest
energy configuration are denoted in \rfig{fig:m_vs_strain} with
respect to the change of lattice parameter $c$.

Only C doping on the B sites noticeably increases the lattice
parameter $c$ and magnetization, while most other substitutions
decrease the magnetization in {\fealb}. Both chemical effect and the
magnetoelastic effect contribute to the magnetization enhancement. The
lattice parameter $c$ increases by \SI{1.6}{\percent} and
magnetization increases to \SI{1.4}{\mu_B\per Fe} in {\fealbc}. Unlike
B, C has a small moment parallel to the Fe sublattice. Moreover, C
doping increases the moments of neighboring Fe atoms by about
\SI{0.1}{\mu_B\per{Fe}}.

For Si doping, without considering lattice relaxation, magnetization
decreases by \SI{6}{\percent} in Fe$_2$(Si$_{0.5}$Al$_{0.5}$)B. The
relaxation decreases the lattice $c$ by \SI{1.7}{\percent}, and
further decreases the magnetization by another \SI{6}{\percent}. Thus,
both chemical effect and the magnetoelastic effect contribute to the
decreasing of the magnetization. Be and Mg dopings have stronger
effects on decreasing the magnetization. With Be and Mg dopings, the
DOS peak right below $\efermi$ in the majority spin shifts toward
$\efermi$ and becomes less occupied and the magnetization decreases to
about \SI{1.1}{\mu_B\per Fe}. Ga doping has very small effect on the
lattice parameters and the magnetization of {\fealb}.

\section{Conclusion}
Using density functional theory, we investigated the intrinsic
magnetic properties in {\tmalb} and their alloys. For {\fealb}, the
$a$ axis is the easiest axis, while the $c$ axis is the hardest
axis. For {\mnalb}, we predict that the magnetic ground state is an
AFM configuration, with the neighboring Mn layers being
antiferromagnetically coupled along the $c$ axis. {\coalb} is weakly
ferromagnetic, while {\cralb} and {\nialb} are nonmagnetic. All $3d$
substitutions decrease the magnetization and Curie temperature of
{\fealb} in the sequence of $\Mn<\Cr<\Co<\Ni$. However, Mn promotes
antiferromagnetism when its doping content is larger than
\SI{20}{\percent}. The competition between the two configurations at
critical compositions may be responsible for the spin-glass states
observed in experiments. Unlike Mn, Cr doping is not likely to promote
the AFM configuration, and may be useful in tuning $\tc$ in
{\fealb}. The effects of strain and alloying on magnetic properties
are also studied. A very strong anisotropic magnetoelastic effect is
found. Magnetization in {\fealb} becomes fragile and quickly decreases
with the lattice parameter $c$, while it barely changes with $a$ and
$b$. This effect is explained by the displacement of antibonding
$d_{3z^2-1}$ states right below the Fermi level in the majority spin
channel, and the filling of unoccupied $d_{xy}$ and $d_{x^2-y^2}$
states which have a sharp peak right above the Fermi level in the
minority spin channel. Doping or applying pressure to modify the
interlayer distance along the $c$ axis may provide an effective way to
tune the magnetic properties in {\fealb}.

\section{Acknowledgments}

We thank R. W. McCallum, L. Lewis, R. Barua, and B. Jensen for helpful
discussions. Work at Ames Laboratory was supported by the
U.S. Department of Energy, Advanced Research Projects Agency-Energy
(ARPA-E) under Grant No. 1002-2147. Ames Laboratory is operated for
the U.S. Department of Energy by Iowa State University under Contract
No. DE-AC02-07CH11358.

\bibliography{aaa,methods}
\bigskip

\end{document}